\documentclass[conference]{IEEEtran}
\IEEEoverridecommandlockouts
\usepackage{cite}
\usepackage{amsmath,amssymb,amsfonts}
\usepackage{algorithmic}
\usepackage{graphicx}
\usepackage{textcomp}
\usepackage{xcolor}

\usepackage[caption=false,font=footnotesize,labelfont=sf,textfont=sf]{subfig}

\def\BibTeX{{\rm B\kern-.05em{\sc i\kern-.025em b}\kern-.08em
    T\kern-.1667em\lower.7ex\hbox{E}\kern-.125emX}}
\begin{document}

\title{A Secure and Efficient Direct Power Load Control Framework Based on Blockchain}

\author{\IEEEauthorblockN{ Ali Dorri}
\IEEEauthorblockA{\textit{CSE, UNSW} \\
		\textit{and   DATA61 CSIRO} \\
Sydney, Australia \\
ali.dorri@unsw.edu.au}
\and
\IEEEauthorblockN{ Fengji Luo}
\IEEEauthorblockA{\textit{University of Sydney}\\
Sydney, Australia \\
Fengji.luo@sydney.edu.au}
\and
\IEEEauthorblockN{ Salil S Kanhere}
\IEEEauthorblockA{\textit{CSE, UNSW} \\
Sydney, Australia \\
Salil.kanhere@unsw.edu.au}
\and
\IEEEauthorblockN{Raja Jurdak}
\IEEEauthorblockA{\textit{DATA61 CSIRO} \\
Brisbane, Australia. \\
Raja.Jurdak@csiro.au}
\and
\IEEEauthorblockN{ Zhao Yang Dong}
\IEEEauthorblockA{\textit{UNSW} \\
	Sydney, Australia. \\
		Joe.dong@unsw.edu.au}
}

\maketitle

\begin{abstract}
	
	Security and privacy in Direct Load Control (DLC) is a fundamental challenge in smart grids. In this  paper, we  propose a  blockchain-based framework to increase security and privacy of DLC. We propose a method whereby participating  nodes share their data with the distribution company in an anonymous and secure manner. To reduce the associated overhead for data dissemination, we propose a hash-based transaction generation method. We also outline the DLC process for managing the load in consumer site. Qualitative  analysis demonstrates the security and privacy of the proposed method. 
\end{abstract}

\section{Introduction}\label{sec:intro}
Power systems are one of the fundamental infrastructures of human societies. The proposal of ``smart grid" in the early 21\textsuperscript{st} century \cite{farhangi2010path} has brought about profound transformations to the operation of modern power systems. Development of smart grid has been recognized as a way to relieve the grand challenges of energy shortage and climate change.
\par 
A key enabler of smart grid technology is the deployment   of Advanced Metering Infrastructure (AMI) \cite{advancedmeter}. A smart meter, which is essentially an electronic device with communication and computing functions, is a central component of AMI and facilitates two way communication between customers and the power utility company. In traditional power systems, the information flow is one-way, which means the utility company collects power demand (known as “load”) information from the customer side, and schedules power generation to serve the load. The growth in the smart grid technology enables  two-way communications  where  the customer can not only send load data to the utility, but also  receive control, electricity tariff, and incentive signals from the utility. Based on the signals, the customer's load profile can be re-shaped to balance load and demand in the grid. This is referred as the ``demand response" or ``Demand Side Management (DSM)" \cite{palensky2011demand} in the literature. \par 

The two-way communication introduces new  security and privacy concerns for the consumers. The attackers may   gain access to the devices in the consumer site by compromising the communications. The load data generated by the consumer can be tracked by malicious nodes which in turn compromises the user privacy.  Thus,  ensuring communication security and privacy is one of the primary considerations in  DSM. \par 

Direct Load Control (DLC) is one of the fundamental DSM methods \cite{chen2014distributed}. In  DLC, DIStribution COmpany (DISCO) centrally collects load and demand requests and decides on re-shaping energy consumption patterns for customers. However, customers cannot verify the action of the energy utility, which for example could include reducing the  energy consumption of the consumer. This is critical as DISCO may maliciously decide on particular actions to increase its profit. DISCO can access and control the devices installed in the user site based on an initially signed contract. However, the user must be able to monitor and check DISCO access to the  devices and  the exchanged data. \par  

Blockchain, a distributed, immutable, secure, and anonymous ledger of blocks, provides a possible solution to address the outlined challenge.  The participating nodes in the blockchain communicate through transactions, which are broadcast and verified by all participating nodes, thus achieving distributed management of blockchain. Each block maintains the hash of the previous block which ensures immutability as modification of a block changes the block hash in the consequent block. Participating nodes use changeable  Public Keys (PKs) as their identity which introduces a level of  anonymity.  \par 

The main contribution of this  paper is to apply blockchain to DLC to protect user privacy and security of communications. Given that all participating nodes must be known to the DISCO, we employ a permissioned blockchain. Thus, only authorized nodes by DISCO can join the blockchain.  The participating nodes periodically send their load/demand data to the DISCO, which in turn increases processing overhead in the participants for generating and signing transactions with asymmetric encryption. To address this challenge, we propose a hash-based demand/load transaction, where the transaction generator is no longer required to sign the transaction. The authentication of the transaction generator and integrity of the data are achieved using a secret hash value which is only known by the transaction generator and DISCO. \par  

Once DISCO collects the load/demand data, it adjusts the energy consumption of participating nodes based on   DLC. Initially, DISCO and the user that wish to participate in DLC sign a contract that contains details of DLC, e.g., the type of sensors that can be installed. Then, DISCO installs sensors in the consumer site to monitor a particular aspect of the site to prevent inconvenience service, e.g., turning off the cooling system during hot weather. As outlined earlier in this section, the consumer must monitor the access of DISCO to his devices and the exchanged data by sensors. In our framework, all communications between participants are stored in the blockchain in the form of transactions. Thus, the consumer can monitor the transactions generated by his devices for  access control. In summary the key contributions of this paper are as follow:  \par 
 \begin{itemize}
 	\item We propose  a blockchain-based solution to DLC that ensures user anonymity and security of communications. 
 	\item Our method reduces the processing overhead for sending demand/load data while ensuring user anonymity  and data integrity.
 	\item We introduce an access control mechanism which allows users to monitor devices and sensors installed by DISCO.
 \end{itemize}

The rest of the paper is organized as follow. Section \ref{sec:introduction} introduces the fundamental concepts of DSM. The proposed blockchain-based DLC is presented in \ref{sec:BDLC}. Section \ref{sec:secuirty} provides a qualitative analyses on security and privacy of the proposed framework. Finally, Section \ref{sec:conclusion} concludes the paper and outlines future work.

\section{Introduction of Demand Side Management}\label{sec:introduction}
This section presents an  introductory discussion of DSM. The goal of DSM is to action certain control decisions such as offering incentives to re-shape load profile of the customers to match the cumulative generation profile.  The DSM actions can be broadly categorized as indirect load control and direct load control and are discussed below. \par 

\vspace{0.2cm}
\textit{A. Indirect Load Control (ILC)}\par 
\vspace{0.2cm}\par 

In ILC,  also known as pricing-based DSM, DISCO utilizes    time-varying electricity pricing signals to stimulate customers to shift their power consumption from peak hours to off-peak hours, which eventually  reduces the   peak load of the grid.   By setting high electricity prices in peak-demand hours and low  prices in off-peak-demand hours, the user is encouraged to shift the use of some appliances such as washing machines to low price hours,  to save the household's  electricity cost.  \par 

The most widely adopted time-varying electricity tariff is  Time-of-Use (TOU) which is widely employed in many countries \cite{de2013time}. In TOU, one day is divided into multiple periods and different electricity prices are applied in each period.    \par 

\vspace{0.2cm}
\textit{B. Direct Load Control (DLC)}
\vspace{0.2cm}\par 

In DLC, also known as incentive-based DSM, DISCO directly asserts control of some customer-side appliances in particular hours, e.g., noon time on some hot summer days  \cite{chen2014distributed}. A contract is signed between the customer and DISCO to agree on the specific hours during which DLC is enforced and the appliances which will be controlled and the actions. Then in the agreed hours, DISCO can control the ON/OFF status or adjust power consumption of the appliances. As a reward, DISCO often provides an economic subsidy to the customer, such as cash reward or electricity price discount. A representative industrial example of DLC is the SmartAC program of Pacific Gas and Electric (PG\&E) \cite{smartprogram}, in which the air conditioners of a group of volunteering residential users are controlled by the utility for  load management. \par 

In summary, although  ILC is unintrusive to the user, it is less reliable for DISCO for performing expected load shifting as there is  no guarantee that the customers shift their  energy usage pattern to respond to the time-varying electricity price. In DLC, however, DISCO can directly control the appliances and thus has the ability to enforce changes.  However, security and privacy concerns remain major challenges in applying DLC as outlined in Section \ref{sec:intro}. In this paper, we propose the use of   blockchain to address these challenges. 

\section{Blockchain-based Direct Load Control }\label{sec:BDLC}

\begin{figure}
	\begin{center}
		\centerline{\includegraphics[width=8cm,height=8cm,keepaspectratio]{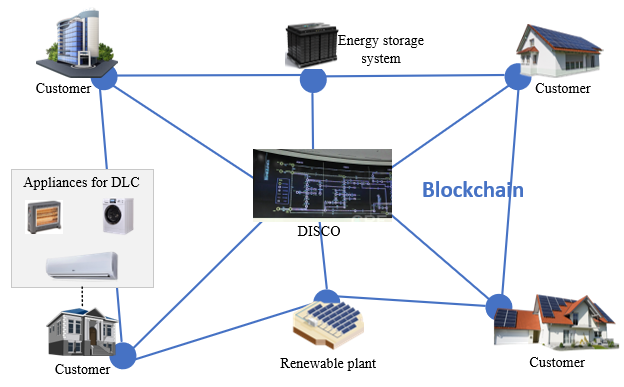}}
		\caption{An overview of the proposed architecture.}
		\label{fig:overall}
	\end{center}
\end{figure}

In this section, we discuss the details of the proposed blockchain-based DLC  framework. The  participating nodes in the system   which include energy producers, consumers, battery energy storage systems, and  DISCO jointly form  a distributed network as shown in Figure  \ref{fig:overall}. The identity of the participating nodes is known to the DISCO.  Thus, we use a permissioned blockchain where participating nodes must be authorized by  DISCO to join the blockchain. DISCO generates a genesis transaction, i.e., the first transaction in each ledger, for each participant which enables them to generate and store transactions in the blockchain.  \par 

The proposed framework consists of two fundamental steps: 1) Data dissemination where the participating nodes share their data with DISCO, and 2) DLC where the DISCO controls the load in the consumer site by adjusting energy consumption of particular devices. The DISCO first has to collect load/demand data from the participants in step 1. Accordingly, DISCO decides on proper load adjusting actions for step 2. To reduce the associated processing overhead on the participants, the structure of the transactions is different in each step. In the first step, where the participants frequently generate transactions,  we propose a hash-based transaction generation method that does not require asymmetric encryption. The second step uses PK and signatures for transaction generation as conventional blockchains. Next, we outline the specific details of these steps.

\subsection{Data Dissemination}
Initially,   DISCO authorizes the  nodes to join the permissioned blockchain.  DISCO needs to know the demand and load of each participating node  to balance the cumulative load with the response and  purchase or sell energy if required. The  participating nodes periodically report their load or demand to the DISCO by  generating and signing transactions using asymmetric encryption. This incurs significant overhead on participating nodes particularly given that  most of participating nodes have limited computation  resources, e.g., smart meters. To  reduce the associated overheads,  our method  proposes a hash-based verification mechanism as outlined below. \par 

To share data, each node generates a \textit{Demand Load (DL)} transaction. The structure of the DL is as below: \par 
$ ID || Data || DLFlag  || Secret $\par 
Where \textit{ID } is the unique identifier of the transaction generator. During the bootstrapping, each node generates  an ID, a pattern, and a secret value and shares with DISCO. To protect anonymity, the node changes the ID for each transaction  using a pattern, e.g., adding previous ID with a constant value. Conceptually, this is similar to changing PK in conventional blockchains. To verify the transactions, DISCO stores all these fields and update them accordingly once it receives transactions (outlined in detail in the rest of this section). \textit{Data} is the amount of demand or load for the user. \textit{DLFlag} identifies whether the value in Data field is demand (DLFlag=0) or load (DLFlag=1). To reduce the overheads on nodes for generating DL translations, a \textit{Secret} is stored in the last field which is the hash of three parameters which are:  \par 
 $H(secret\_value || nonce || Data)$\par 
Where \textit{H(x)} is hash of x. The  \textit{secret\_value}  is generated by the node during bootstrapping.  \textit{Nonce} is a one-time value used to prevent replay attack which conceptually is similar to the  nonce   employed in   Ethereum transactions \cite{wood2014ethereum}. The \textit{Data} value is  included  to prevent any node from changing the Data. \par 
Each node generates and broadcasts the DL transactions in pre-defined time intervals.  Once received, the DISCO extracts the ID of the transaction generator and  locates the stored details (that are secret\_value, nonce, and pattern) for that ID. Then, the DISCO generates Secret, as outlined above, with received Data and the corresponding nonce and secret\_value in its own record.  If the resulting hash matches with the Secret, the transaction is verified. The DISCO then updates nonce and ID for the node for future transactions. If the hash does not match with the Secret, the transaction is dropped. \par 

DISCO creates a Merkle tree of all received DL transactions in a period. Next, it stores the signed  root hash of the Merkle tree in the blockchain. Storing only the root hash of the Merkle tree in place of all received DL transactions reduces the processing overhead  and memory requirement for managing the blockchain and thus increases its scalability. Note that the demand/load data are only critical for the DISCO and other participants  are typically only required  to prove the generation of their transactions  which can be done using the root hash of the Merkle tree. \par

\subsection{Direct Load Control (DLC)}
\begin{figure}[h]
	\begin{center}
		\centerline{\includegraphics[width=9cm,height=8cm,keepaspectratio]{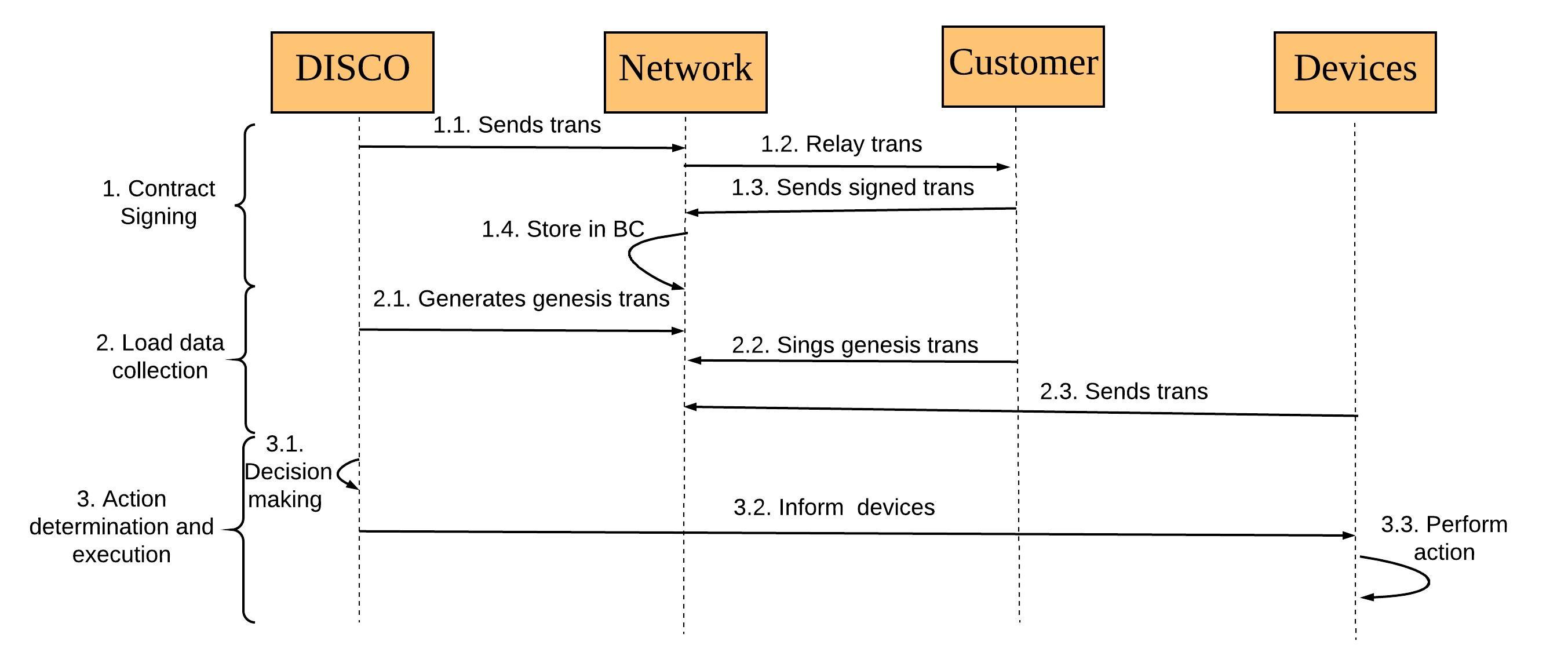}}
		\caption{Illustration of the transactions exchanged during the DLC process.}
		\label{fig:diagram}
	\end{center}
\end{figure}

In DLC, the DISCO manipulates the usage of devices on the customer premise to manage the load, based on the received DL transactions.  DISCO and the customer would typically sign a contract that allows DISCO to control energy usage in the customer site. The DISCO then installs sensors at the customer site to measure real-time data to prevent any service inconvenience. This process is summarized in Figure \ref{fig:diagram} and is illustrated in details below.  \par 

\textit{(1) Contract signing}:  During the bootstrapping the load control  contract is signed by the DISCO and the customer which is used to manifest DLC. The contract contains the type of devices in the consumer site to be controlled by the DISCO, the type and number of sensors that DISCO can install, and the duration when the DISCO can change the energy consumption in the consumer site.  DISCO initiates the contract signing process by generating and signing a\textit{ load control} transaction that is structured as below: \par 
$T\_ID || P\_T\_ID || PKGen || SignGen  || PKRec || SignRec  \\ || Ref.DISCO.ID  || Metadata$\par 
Where T\_ID is the transaction ID which is essentially the hash of the transaction content.  DISCO initially generates genesis transaction for each authorized customer. The customer chains all its future transactions to the genesis transaction by storing  the ID of  his previous transaction in P\_T\_ID field of new transactions. The load control transaction is a multisign transaction meaning that it requires signature of two entities  to be stored in the blockchain.  The next four fields are   the PK and  signature of the transaction generator and receiver. The  sensors installed in the customer site may generate transactions in response to requests from  DISCO, e.g., DISCO may request temperature sensors in the consumer site to periodically send site temperature to it. The verification of these transactions involves verifying  the corresponding request transaction generated by the DISCO. This ensures that the  actions performed by devices or sensors, e.g., turning off the cooling system, is aligned with the DISCO request. Thus, the  participating nodes  have to search the entire ledger  of  DISCO to find the corresponding request transaction, which  incurs significant latency and processing overheads.  To  reduce this overhead, the ID of the DISCO request transaction that has triggered the generation of the current response transaction is stored in Ref.DISCO.ID field.  The latter  applies for transactions generated by the customer only. For DISCO transactions, this field is set to NULL.  Metadata field includes extra information that the two communication entities may wish to exchange. \par 

The contract between DISCO and the customer is  encrypted with the PK of the customer, to ensure confidentiality, and is put in the metadata of the transaction. Next, the transaction is broadcast (1.1 Fig \ref{fig:diagram}).    On receipt (1.2), the customer decrypts the contract using its PK to verify the contract content. If the customer agrees with the contract, he proceeds to  sign the received  transaction by populating SignRec  and broadcasts it to the blockchain (1.3).  The miners then store the transaction in the blockchain as it contains both required signatures. \par  

\textit{(2) Load data collection}: Based on the contract, the DISCO installs   sensors in the customer site for monitoring purpose. Initially, DISCO  creates a   genesis transaction for each sensor   which contains a virtual link to the  transaction that contains the contract (using Ref.DISCO.ID field) (2.1). The sensors collect and share particular data from the consumer site depending on sensor type, e.g., temperature. The genesis transaction needs to be signed by both the DISCO and the customer to ensure that installed sensors are as defined in the contract  (2.2). Once deployed, the sensors periodically send data to  DISCO using DL transactions (2.3). \par
\textit{(3) Action determination and execution}: Based on the load data collected from sensors,  DISCO determines the load control actions (3.1) using methods such as \cite{ruiz2009direct}.  DISCO then  informs the devices in the consumer site  of the identified action using a load control transaction (3.2). On receiving the latter, the devices first verify the transaction by validating the signature with the PK of the transaction. Once verified, the devices   perform the indicated action in the transaction (3.3). \par

The outlined process enables the DISCO to mange load in the grid in a secure and private manner. 

\section{Security and Privacy Analysis} \label{sec:secuirty} \label{sec:analys}
In this section, we analyze the security and privacy of the proposed method.\par 
\textit{Security:} In the proposed method, the transactions contain the signed hash of the exchanged data which achieves integrity of data and non-reputability. Recall that in DL, transactions are generated without any asymmetric encryption and  rely on a hash value for authentication, data integrity, and non-repudiation. Recall from Section \ref{sec:BDLC} that the ``secret" in the DL transaction   is the hash of a secret\_value along with data and a nonce. The secret\_value can be utilized to authenticate the consumer as the secret is only known by the consumer and DISCO. The Data field ensures integrity of the data as if  a malicious node modifies data, the Secret generated by DISCO will not match with the Secret in the received transaction. Nonce  protects against replay attack wherein a malicious node repeats a communication by storing  a transaction locally and re-sending it to the network at any point of time. DISCO discards transactions with duplicate nonces.  \par 

\textit{Privacy:}  The privacy of the proposed method is inherited from the anonymity of the blockchain. Recall from Section \ref{sec:BDLC} that each participant uses a changeable PK as its identity that introduces a level of anonymity among entities other than DISCO. This protects against malicious nodes that may track a costumer using its PK. Similar to other existing blockchain instantiations, linking multiple PKs in the blockchain to deanonymize a user, known as a linking attack in literature,  is possible in the proposed method. \par 
The DL transaction contains  the load and demand of the consumer which constitutes privacy-critical information. The proposed framework achieves a level of  anonymity  for DL transactions as: i) the transactions are generated independently, i.e., without relying on a previous transaction, and ii) only the root hash of a Merkle tree of all DL transactions is stored in the blockchain, thus there is less information for the attackers to conduct linking attack. \par 
 The consumer  can monitor the identity of the nodes that have accessed its sensors, the frequency and duration of access, and the shared information as all transactions are recorded in the blockchain and can be monitored. \par

\section{Conclusion} \label{sec:conclusion}

In this  paper, we proposed a blockchain-based direct load control framework.  We discussed  disseminating  the load data from distribution nodes to the distribution company  as well as the direct load control process.   To reduce the associated processing overhead on participating nodes for frequently sharing demand and load data, we proposed a hash-based demand load transaction. Qualitative security and privacy analysis demonstrate the security and anonymity of the proposed method. \par 

\bibliographystyle{IEEEtran}
\bibliography{bare_jrnl_compsoc}

\end{document}